\documentclass[printer]{aa}
\usepackage{amssymb}
\usepackage{amsmath}
\usepackage{txfonts}
\usepackage{graphicx}
\usepackage{natbib}
\usepackage{rotating}
\usepackage{url}
\usepackage{epsfig}
\usepackage{longtable}

\def\gr{$\gamma$-ray}

\usepackage{color}
\definecolor{red}{rgb}{0.7,0,0}
\definecolor{blue}{rgb}{0,0,0.7}

\begin{document}

\title{Search for variable gamma-ray emission from the Galactic  plane
in the Fermi data.}
    \author{A. Neronov \inst{1}, D.Malyshev\inst{2}, M.Chernyakova\inst{3},
          A.Lutovinov\inst{4}
          }

   \institute{ISDC Data Centre for Astrophysics, Ch. d'Ecogia 16, 1290, Versoix, Switzerland \\  \and Bogolyubov Institute for Theoretical Physics, Metrologichna str., 14-b, Kiev 03680, Ukraine \\ \and School of Physical Sciences, Dublin City University, Glasnevin, Dublin 9, Ireland ; DIAS, Fitzwiliam Place 31, Dublin 2, Ireland \\
         \and Space Research Institute, Profsoyuznaya 84/32,117997 Moscow, Russia
             }

\titlerunning{Variable gamma-ray emission from the Galactic  plane}

\abstract
{High-energy gamma-ray emission from the Galactic plane above
$\sim$ 100 MeV is composed of three main contributions: diffuse emission from cosmic ray interactions in the interstellar medium, emission from extended sources, such as supernova remnants and pulsar wind nebulae, and emission from isolated compact source populations.}
{The diffuse emission and emission from the extended sources provide the dominant contribution to the flux almost everywhere in the inner Galaxy, preventing the detection of isolated compact sources. In spite of this difficulty, compact sources in the Galactic plane can be singled out  based on the variability properties of their \gr\ emission. Our aim is to find sources in the Fermi data that show long-term variability. }
{We performed a systematic study of the emission variability from the Galactic plane, by constructing the variability maps.}
{We find that emission from several directions along the Galactic plane is significantly variable on a time scale of months. These directions include, in addition to known variable Galactic sources and background blazars,  the Galactic ridge region at positive Galactic longitudes and several regions containing young pulsars. We argue that variability on the time scale of months may be common to pulsars, originating from the inner parts of pulsar wind nebulae, similarly to what is observed in the Crab pulsar.  }
{}

\maketitle

\section{Introduction}

Most of the \gr\ emission from the Galactic plane in the energy band above 0.1~GeV is produced by cosmic ray interactions with interstellar matter \citep{fichtel75,hunter97, fermi_diffuse}. These interactions, on a time scale of $\sim 10^7$~yr and on distance scales of hundreds of parsecs, result in the production of bright large-scale diffuse emission from the entire Galaxy. Isolated Galactic \gr\ sources, such as pulsars, pulsar wind nebulae, supernova remnants, and \gr -loud binary systems are superimposed on this large-scale diffuse emission. The strong diffuse background emission provides the main obstacle for detecting individual isolated sources in the Galactic plane. Inhomogeneities of the diffuse emission induced by the inhomogeneity of matter distribution in the interstellar medium lead to  local brightness enhancements of the \gr\ emission from the Galactic plane, which might be misinterpreted as isolated sources. Some of the sources at low Galactic latitude, reported in the two-year Fermi catalog, might have such a nature \citep{fermi_catalog}.

\gr -loud binaries provide a very moderate contribution to the overall \gr\ emission from the Galaxy. Up to now, only a few binary systems were found to produce \gr s with energies above 0.1~GeV. Six out of nine known \gr -loud binaries, PSR B1259-63 \citep{b1259}, LSI +61 303 \citep{lsi}, LS 5039 \citep{ls}, Cyg X-3 \citep{cygx3}, Cyg X-1 \citep{cygx1}, and   HESS J0632+057 \citep{hessJ0632} are systems composed of a compact object (a neutron star or a black hole) orbiting a massive companion star. One system, Eta Car \citep{etacar}, is a massive star binary with Wolf-Rayet star, one system, V407 Cyg \citep{V407Cyg_fermi}, is a symbiotic binary containing a white dwarf accreting matter from the main sequence star. The nature of the recently discovered source 1FGL J1018.6-5856 is not well constrained yet  \citep{1FGLJ1018_fermi}.

Indeed, only a very small fraction of the known high-mass binaries with a compact object orbiting a massive star (most of them are discovered as high-mass X-ray binaries, HMXRB) turn out to be ``\gr -loud''. It is not clear at the moment if these are only few detected \gr -loud binary systems because only a small fraction of the HMXRBs are able to produce \gr s or because most of the HMXRBs are emitting \gr s, but their typical flux levels are just below the sensitivity of Fermi. It is also not clear which properties of the HMXRB are responsible for the enhanced \gr\ emission from the four known \gr -loud HMXRBs.  The distinguishing feature of the \gr -loud binary systems is variability. Other known types of Galactic \gr\ sources, including the diffuse emission and supernova remnants, are not variable on day-to-month time scales. This feature could reveal the presence of \gr\ -loud binaries even at the top of much stronger diffuse emission.

Among pulsars, only the Crab pulsar, which is the youngest radio pulsar in the Galaxy, is known to produce \gr\ flares on time scale of hours-to-days \citep{crab_flares}. The variable emission is produced in the inner part of the pulsar wind nebula (PWN) associated to the Crab pulsar. It is not clear if the flaring activity of the nebula is a peculiar feature of the Crab pulsar or if it is a generic property of the young-pulsar-powered systems. A systematic study of  variability properties of other known young pulsar systems in the Galaxy is needed to clarify this question.

Fermi/LAT is a pair-conversion gamma-ray detector operating between 20
MeV and 300 GeV. The LAT has a wide field of view of 
sr at 1 GeV, and observes the entire sky every two orbits \citep{Atwood09}.
Therefore it is a perfect telescope for the systematical study of the  variability of the GeV sky on a scale of hours-to-days.


\section{Search for variability of \gr\ emission in the Galactic plane}

A simple way to verify if \gr\ emission from a given direction on the sky is variable or constant is to analyze the lightcurve and check if it is consistent with a constant flux. Several measures of (in)consistency with the constant flux can be considered.

Consider the flux $F(l,b,t_i)=F_i(l,b)$ in the given sky direction $(l,b)$ (in Galactic coordinates) binned in $N$ time bins $1\le i\le N$. The flux is consistent with constant if the reduced $\chi^2$ of the fit of the lightcurve with the constant
\begin{equation}
\chi^2(l,b)=\frac{1}{N-1}\sum_{i=1}^N \frac{(F_i-\overline F(l,b))^2}{\sigma_i^2}
\label{chi2}
\end{equation}
is close to one (here $\sigma_i$ are the measurement errors in each time bin and $\overline F(l,b)$ is the average flux in the $(l,b)$ direction). If we consider the dependence of $\chi^2$ on $(l,b)$ we expect to find $\chi^2\simeq 1$ in all directions except for the directions in which there are variable isolated \gr\ sources. All non-variable \gr\ sources, such as supernova remnants and the large-scale diffuse Galactic emission should not appear in the $\chi^2(l,b)$ map. Instead, the sources that are characterized by a variability amplitude larger than $\sim \sigma_i/\overline F(l,b)$ should be visible as excess in the\ $\chi^2$ map.

The scatter of the source flux around its average value, characterized by the $\chi^2$ of the fit of the lightcurve with a constant, provides a good measure of variability especially for sources that are systematically variable throughout the observation period. On the other hand, sources that produce on average one flare with a duration much shorter than the span of the observation period are gradually "washed out" from the $\chi^2(l,b)$ map as the overall exposure time increases. Indeed, the contribution of the time bins containing the flare to the overall $\chi^2$ given by Eq. (\ref{chi2}) decreases with increasing number of bins. Accordingly, single-flare sources might disappear from the $\chi^2$ map with large exposure. A better measure of variability for the single-flare sources is the maximum deviation of the flux from the average value
\begin{equation}
\sigma_{max}(l,b)=\mbox{max}_{i=1}^N\left(\sqrt{\frac{(F_i-\overline F(l,b))^2}{\sigma_i^2}}\right).
\end{equation}
Similarly to the $\chi^2$ measure, excess of $\sigma_{max}$ beyond the expected statistical fluctuations  indicates the presence of a flaring source in a given sky direction. In contrast to the $\chi^2$ measure, the $\sigma_{max}$ measure does not decrease with increasing exposure.

A systematic view on the entire population of variable sources in the Galaxy (both systematically variable and flaring) can be obtained by identifying all significant excesses in the $\chi^2 (l,b)$ and $\sigma_{max}(l,b)$ maps constructed based on the \gr\ lightcurves in different energy bands and different energy scales. In the following sections we investigate these sets of "variability maps" of the \gr\ emission from the Galaxy.

Collecting the photons within a PSF (point spread function) can lead to false-positive or misaligned detections of variability. Consider the case of $N$ strongly variable sources with maximum flux deviations from constant level at $\sigma_i$, $i=1..N$. These sources can cause an excess of variability at some region, at which the contributions from the PSF of all these sources are non-negligible. 
Namely, if $\mathcal{P}(r)$ is the PSF (the fraction of source's photons beyond radius $r$) of Fermi/LAT and $\alpha_i$ is the angle under which the region is visible from $i$-th variable source position, it can be shown that the significance of false-variability of this region can be estimated as
\begin{equation}
\sigma_{false}\leq\sum\limits_{i=1}^{N}\sqrt{\frac{\mathcal{P}(r_i)\alpha_i}{2\pi}}\sigma_i,
\label{sigma_false}
\end{equation}
where $r_i$ is the angular distance from $i$-th variable source to the selected region.

Consider now the case of a flaring weak source in the vicinity of bright non-variable source. In this case the non-variable bright source will contribute  ``background'' photons at the position of the weak source. This background reduces the significance of the flare in photons collected within the PSF exactly around the position of the flare. However, the number of ``background'' photons will decrease with increasing distance from the bright non-variable source. This can shift the observed maximum of the variability from the flare position along the line connecting a bright and a flaring source in the direction farther from the non-variable source.

Possible false detections of variability caused by the influence of nearby sources are best addressed using variability maps: the imaging of the variability properties of emission helps to spot these possible false detections of variability of non-variable sources  in the presence of known nearby variable sources.

\section{Observations and data analysis}

In our analysis we used all publicly available LAT data for the period from August, 2008 to October, 2011. We processed the data using the Fermi Science Tools software package, version v9r23p1 \footnote{\tt http://fermi.gsfc.nasa.gov/ssc/data/analysis/}.  We filtered the event lists using {\it gtselect} tool and left only "source" events ({\tt evclass=3}) at zenith angles $\theta_z<100$.

To produce the variability maps, we used the "aperture photometry" method. We binned the events from circles of radius $2^\circ$ around each reference position on the sky, using the {\it gtbin} tool. The exposure in each time bin and each sky direction was calculated using the {\it gtexposure} tool and the instrument response function set for the eventclass {\tt P7SOURCE\_V6}.

\section{Results}

\subsection{Variability of emission from the inner part of the Galactic disk}

\begin{figure}
\includegraphics[width=0.49\linewidth,angle=0]{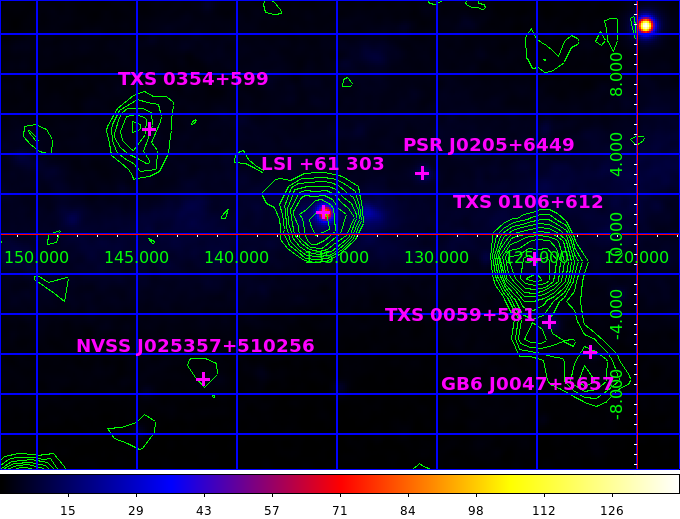}
\includegraphics[width=0.49\linewidth,angle=0]{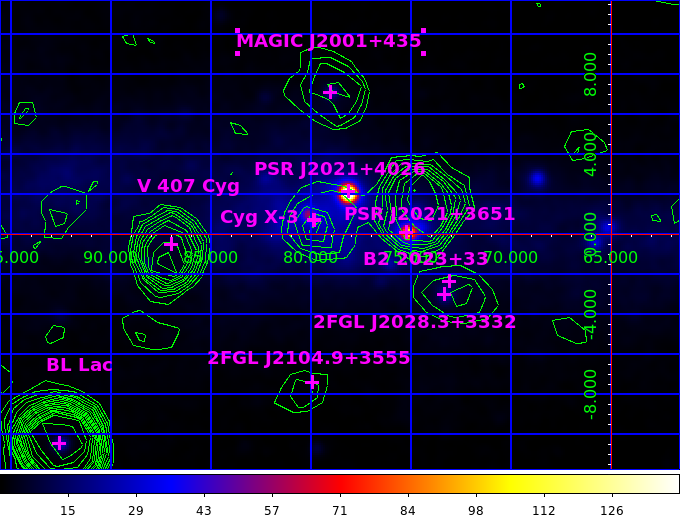}
\caption{LAT count map and variability contours ($\sigma_{max}$ values) in the energy band above 300 MeV in the directions of  the sky regions containing the LSI +61 303 \gr -loud binary (left)  and the Cyg X-3 and V407 Cyg \gr -loud binaries (right). Contour levels start from $3\sigma$ with steps of $0.5\sigma$  }
\label{fig:cygx3}
\end{figure}

The  lightcurves with month-long time bins  for the 3.1~yr exposure of LAT have  37 time bins. One can estimate the probability for $\sigma_{max}$ in any individual lightcurve to exceed a certain threshold value $\sigma_{thr}$. For example, the chance probability for $\sigma_{max}$ to exceed the threshold of $\sigma_{thr}=3.2$ is 5\%, while only 0.2\% of the non-variable source lightcurves would occasionally have $\sigma_{max}$ in excess of $\sigma_{thr}=4$. Producing the variability map of the entire Galactic plane, we have scanned over all different directions along the $360^\circ$ of the plane projection on the sky. The PSF of LAT at energies above 300~MeV has a half-width (68\% contrainment radius) of $2^\circ$. This implies that $\simeq 360^\circ/4^\circ=90$ independent lightcurves were probed. If the flux from the entire Galactic plane is constant, one expects to find less than one lightcurve (on average) in which $\sigma_{max}$ is in excess of $\sigma_{thr}=4$. Consequently, all the directions in which variability in excess of $\sigma_{thr}=4$ is found should be considered as directions from  which real variable emission is detected.

Fig. \ref{fig:cygx3} shows the LAT count map in the energy range above 300~MeV with the superimposed contours showing the $\sigma_{max}$ values found in the lightcurves binned into the month-long time bins for the sky regions containing the known variable Galactic \gr\ sources LSI +61 303, Cyg X-3 and V407 Cyg. All three sources are \gr -loud binaries with quite different types of variability. LSI +61 303 is a persistent source, variable on 26.4~d orbital period and on a longer superorbital period of 4.6~yr. Cyg X-3 is a recurrent flaring source that already exhibited  several bright flares over the 3.5~yr period of Fermi/LAT exposure. V407 Cyg is a symbiotic binary that produced only one flare during the LAT exposure. One can see that in spite of the large difference in the variability properties of the sources, all three sources are clearly detected as strong excesses in the variability map, shown by the green contours. The other excesses above and below the Galactic plane are produced by known blazars.

\begin{figure}
\includegraphics[width=\linewidth,angle=0]{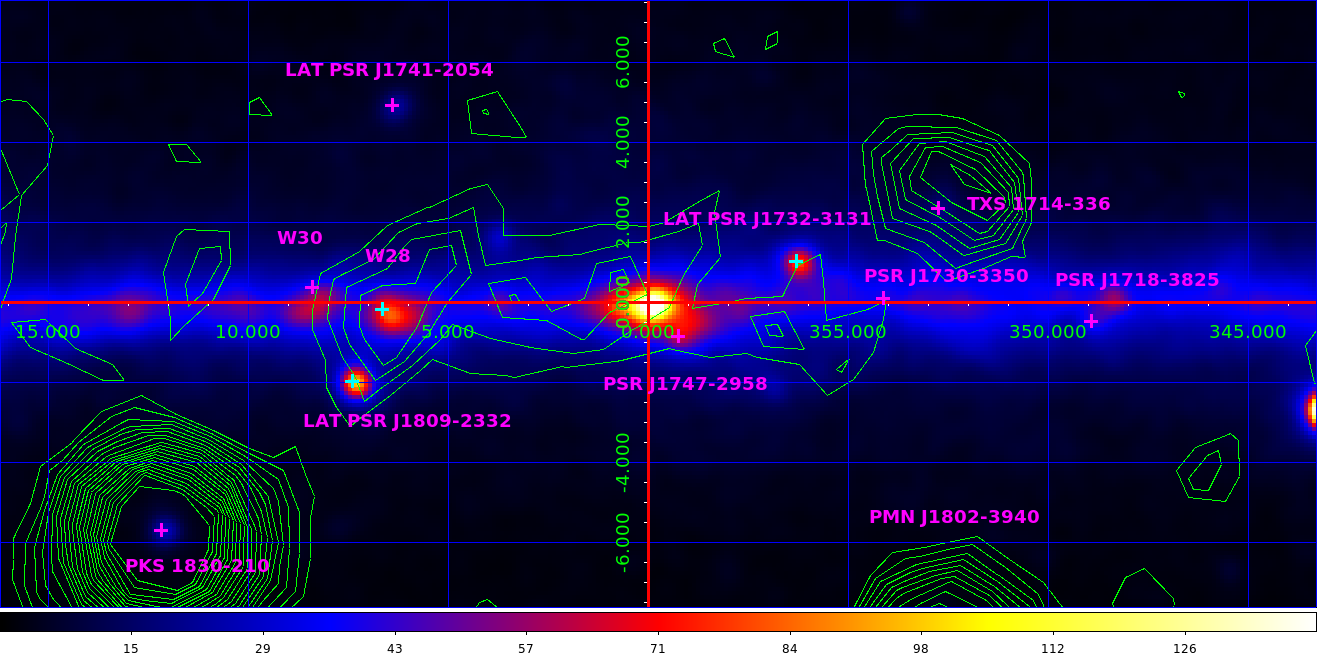}
\caption{LAT count map and variability contours in the direction of Galactic ridge and W28 supernova remnant. Notations are the same as in Fig. \ref{fig:cygx3}. }
\label{fig:ridge}
\end{figure}

Fig. \ref{fig:ridge} shows the LAT count map with superimposed variability contours for the region around the Galactic Center.   The excess variability is not centered at the Galactic Center itself (the bright source in the center of the image). Instead it traces the part of the Galactic plane at positive Galactic longitude $0^\circ<l<7^\circ$ dominated by  diffuse emission, with no clearly isolated bright sources. The end part of the variable region, with the brightest excess at the level of $\sigma_{max}\simeq 4.5$, is centered at the bright isolated source identified with  supernova remnant W28. Appearance of excess variability from the region of diffuse emission and from a supernova remnant is surprising.

\begin{figure}
\includegraphics[width=\linewidth,angle=0]{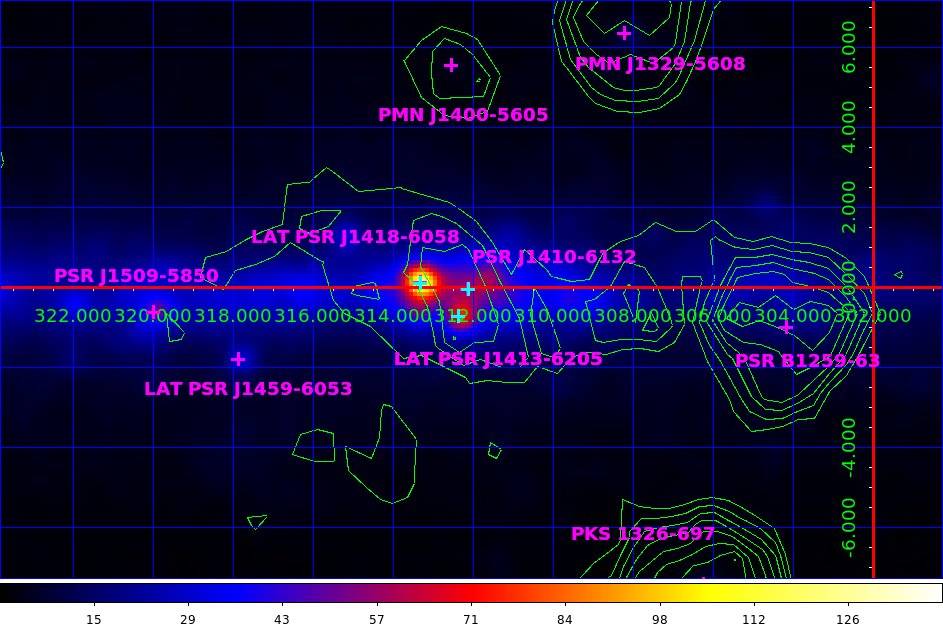}
\caption{LAT count map and variability contours in the direction of  sky region containing pulsars PSR J1418-6068, PSR J1410-6132 and PSR J1413-6205.  Notations are the same as in Fig. \ref{fig:ridge}. }
\label{fig:kookaburra}
\end{figure}

Another region with excess variability is found in the direction of the Kookaburra pulsar wind nebula complex, shown in Fig. \ref{fig:kookaburra}. This sky region contains three young pulsars, PSR J1418-6068, PSR J1410-6132 and PSR J1413-6205, which emit pulsed \gr\ emission in the  0.1-10~GeV band. Similarly to the regions of the Galactic ridge and of LSI +61 303, Cyg X-3 / V407 Cyg, the variable sources above and below the Galactic plane, visible on the map, are background blazars. The source to the right of the Kookaburra complex, PSR B1259-63, is not visible in the three-year count map (shown by the color scale), but appears as a very bright excess in the variability map. This is not surprising, because this source is a \gr -loud binary with a radio pulsar on a highly eccentric orbit with a 3.5~yr orbital period. It exhibits periods of activity only close to the periastron passage. The last periastron passage happened in December 2010. The source produced a bright one-month-long flare in January 2011 and disappeared since then.

Excess  of variability on a scale of months at the level $\sigma_{max}\simeq 15$ is also detected from the direction of the Crab pulsar.

\section{Discussion}

Emission variability from the Galactic plane is caused by compact \gr\ sources. The only firmly established class of generically variable Galactic sources is that of \gr -loud binaries. In this respect, it is not surprising that the highest excess variability is found in the directions of known  \gr -loud binary systems: LSI +61 303, V407 Cyg, Cyg X-3, and PSR B1259-63.

The recent discovery of \gr\ flaring activity of the Crab pulsar shows that pulsars (or the inner compact parts of the pulsar wind nebulae, PWNe) should also be considered as a possible variable Galactic source population. However, it is difficult to draw definitive conclusions about the variability of the entire pulsar population based on the study of a single source, especially taking into account the fact that the Crab pulsar could be considered as a special case because it is the youngest known \gr -loud pulsar.  Our study shows, however, that long (month) time scale variability may indeed be a generic feature of pulsars/PWN systems, as discussed below.

In particular, significant variability is found in the direction of the region of the Kookaburra nebula, which contains several young pulsars. Unfortunately, the point spread function of LAT in the energy band in which the variability is detected is too large 
to tell if one of the pulsars produces  variable emission on a scale of months, or if the variability is the result of the variability of several pulsars. It is also not clear why pulsars in the Kookaburra region would be variable while other young \gr -loud pulsars in the Galactic plane do not reveal any excess variability a time scale of months. Additional study is needed to clarify the origin of \gr\ emission from young pulsars, including the youngest Crab pulsar.

Another region that exhibits excess variability is the Galactic ridge at positive Galactic longitudes $0^\circ<l<7^\circ$.
The strongest excess of variability is centered at the supernova remnant W28. Clearly, the variability in this sky direction cannot be produced by the extended emission from  parts of W28 itself, which have extension of about $0.5^\circ$ \citep{w28_fermi}, 
i.e. $\sim 20$~pc at $\sim 2$~kpc.  The variable region contains several young pulsars, including PSR J1747-2809 and PSR  J1746-2850, close to the Galactic Center  with the ages $T=5.3$~kyr and $T=12.7$~kyr, just slightly older than those of the Crab pulsar. Closer to the W28 remnant there is PSR B1757-24, a pulsar of age $T=15.5$~kyr. Finally, W28 itself was supposed to be associated with the pulsar PSR 1758-23 \citep{w28_pulsar}. It is possible that the variability is produced by one or several pulsars in this region of the Galaxy. Otherwise, the source of variability could be \gr -loud binaries that are below the sensitivity of LAT.

We also remark that the position of this variable region is consistent with the positions of sources that are variable in hard X-ray band (see e.g. \citet{Integral_var}).

The threshold value $\sigma_{thr}$ for significant detection of variability is different for the search of variable emission from an arbitrary direction along the Galactic plane and for the search of variability of emission from known isolated \gr\ source classes. If we are interested in the variability properties of known sources, the threshold $\sigma_{thr}$ can be decreased, because in this case there is no trial factor associated to the scan of the entire Galactic plane. The probability to find an excess variability at the level above $3.5\sigma$ is just about 1\% for each individual source.

\begin{figure}
\includegraphics[width=0.49\linewidth,angle=0]{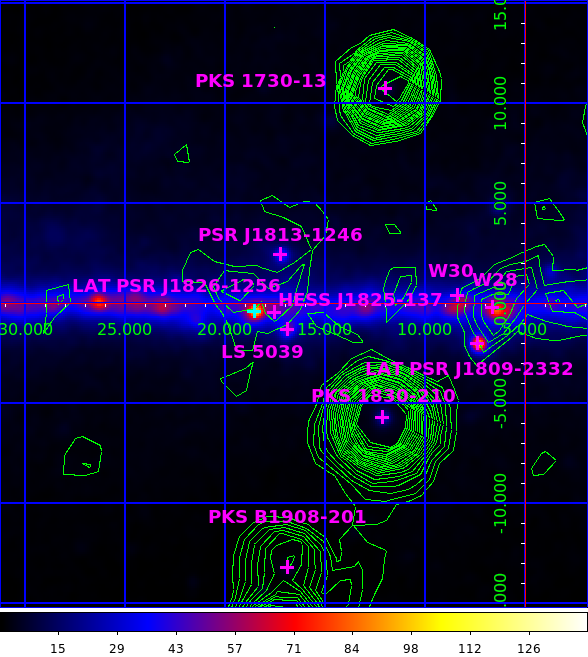}
\includegraphics[width=0.49\linewidth,angle=0]{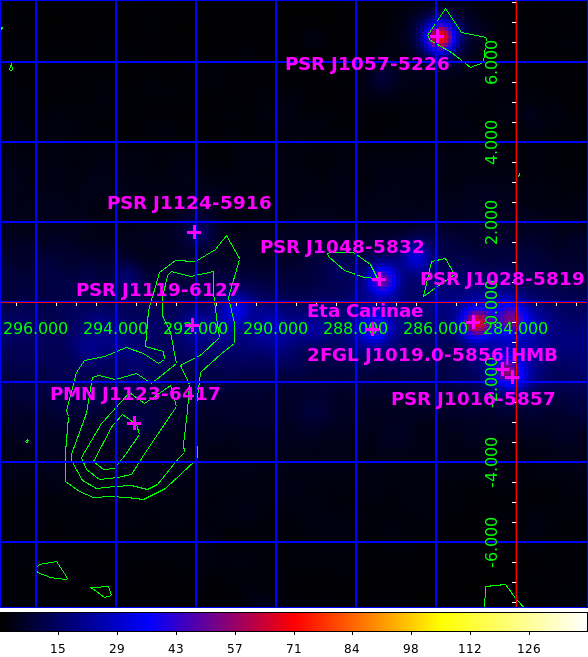}
\caption{Variability contours superimposed on the LAT count map for the sky region of HESS J1825-137. Notations are the same as in Fig. \ref{fig:cygx3}.  }
\label{fig:1825}
\end{figure}

Excess variability at a level higher than $3.5\sigma$ is found in two more directions on the sky, which coincide with the positions of the known sources,  PSR J1826-1256 and PSR J1119-6127. The LAT count maps with the superimposed variability contours in these directions are shown in Fig. \ref{fig:1825}. PSR J1826-1256 is situated close to a known \gr -loud binary LS 5039, which exhibits emission variable on the 4.6~d orbital period. Surprisingly, the emission from LS 5039 appears to be very stable on a time scale of months, so that no excess variability associated with LS 5039 is found. No variability is detected either from the two other binary systems, Eta Car and 2FGL J1019.0-5856 (see right panel of Fig. \ref{fig:1825}). 
Evidence for excess variability at the locations of these pulsars, if combined with the detection of variability from the direction of Kookaburra region that contains several young pulsars, shows that long time scale variability may be a generic property of young pulsars, and not a peculiar feature of the Crab pulsar alone.

To summarize, we have studied variability  of \gr\ emission on a time scale of months from the Galactic plane in the energy band above 300~MeV in the Fermi/LAT data and found several directions with significantly variable emission. We found that strong variability excess is detected in the directions of known \gr -loud binary systems. Remarkably, there are no signatures of weaker variability, which could be clearly associated to other binary systems with black holes and/or neutron stars. At the same time, our study revealed weaker variability excess in the directions of pulsars. This indicates that long variability on a scale of months, as recently discovered in the Crab pulsar, may be a generic feature of the \gr\ emission from the young pulsar population, so that pulsar/PWN systems should be considered as variable Galactic \gr\ sources.\\
\textbf{Acknowledgements.} The authors thank the participants of the ISSI team ``Study of Gamma-ray Loud Binary Systems'' for useful discussions, and the International Space Science Institute (ISSI, Bern) for support. The authors also wish to acknowledge the SFI/HEA Irish Centre for High-End Computing (ICHEC) for providing  computational facilities and support. The work of D.M. is supported in part by the cosmomicrophysics program
of the National Academy of Sciences of the Ukraine and by the State Program of Implementation of Grid Technology in the Ukraine. AL acknowledges the support from the Russian Foundation of Basic Research (grants
11-02-12285-ofim-2011, 12-02-01265), program "Non-stationary phenomena in
objects of the Universe", grant NSh-5603.2012.2 and State contract
14.740.11.0611

\def\aj{AJ}%
\def\actaa{Acta Astron.}%
\def\araa{ARA\&A}%
\def\apj{ApJ}%
\def\apjl{ApJ}%
\def\apjs{ApJS}%
\def\ao{Appl.~Opt.}%
\def\apss{Ap\&SS}%
\def\aap{A\&A}%
\def\aapr{A\&A~Rev.}%
\def\aaps{A\&AS}%
\def\azh{AZh}%
\def\baas{BAAS}%
\def\bac{Bull. astr. Inst. Czechosl.}%
\def\caa{Chinese Astron. Astrophys.}%
\def\cjaa{Chinese J. Astron. Astrophys.}%
\def\icarus{Icarus}%
\def\jcap{J. Cosmology Astropart. Phys.}%
\def\jrasc{JRASC}%
\def\mnras{MNRAS}%
\def\memras{MmRAS}%
\def\na{New A}%
\def\nar{New A Rev.}%
\def\pasa{PASA}%
\def\pra{Phys.~Rev.~A}%
\def\prb{Phys.~Rev.~B}%
\def\prc{Phys.~Rev.~C}%
\def\prd{Phys.~Rev.~D}%
\def\pre{Phys.~Rev.~E}%
\def\prl{Phys.~Rev.~Lett.}%
\def\pasp{PASP}%
\def\pasj{PASJ}%
\def\qjras{QJRAS}%
\def\rmxaa{Rev. Mexicana Astron. Astrofis.}%
\def\skytel{S\&T}%
\def\solphys{Sol.~Phys.}%
\def\sovast{Soviet~Ast.}%
\def\ssr{Space~Sci.~Rev.}%
\def\zap{ZAp}%
\def\nat{Nature}%
\def\iaucirc{IAU~Circ.}%
\def\aplett{Astrophys.~Lett.}%
\def\apspr{Astrophys.~Space~Phys.~Res.}%
\def\bain{Bull.~Astron.~Inst.~Netherlands}%
\def\fcp{Fund.~Cosmic~Phys.}%
\def\gca{Geochim.~Cosmochim.~Acta}%
\def\grl{Geophys.~Res.~Lett.}%
\def\jcp{J.~Chem.~Phys.}%
\def\jgr{J.~Geophys.~Res.}%
\def\jqsrt{J.~Quant.~Spec.~Radiat.~Transf.}%
\def\memsai{Mem.~Soc.~Astron.~Italiana}%
\def\nphysa{Nucl.~Phys.~A}%
\def\physrep{Phys.~Rep.}%
\def\physscr{Phys.~Scr}%
\def\planss{Planet.~Space~Sci.}%
\def\procspie{Proc.~SPIE}%
\let\astap=\aap
\let\apjlett=\apjl
\let\apjsupp=\apjs
\let\applopt=\ao
\bibliography{biblio}
 \label{lastpage}

\end{document}